\documentclass[prb,showpacs,twocolumn,aps,superscriptaddress,a4paper]{revtex4-1}
\usepackage{dcolumn,amssymb,amsmath,amsfonts,graphicx,latexsym,color}
\usepackage{epstopdf}
\begin{document}

\title{Fate of topological states and mobility edges in one-dimensional slowly varying incommensurate potentials}

\author{Tong Liu}
\affiliation{Department of Physics, Southeast University, Nanjing 211189, China}
\author{Hai-Yang Yan}
\affiliation{Key Laboratory of Neutron Physics, Institute of Nuclear Physics and Chemistry, CAEP, Mianyang, Sichuan 621900,China}
\author{Hao Guo$^*$}
\affiliation{Department of Physics, Southeast University, Nanjing 211189, China}
\email{guohao.ph@seu.edu.cn}

\begin{abstract}
We investigate the interplay between disorder and superconducting pairing for a
one-dimensional $p$-wave superconductor subject to slowly varying incommensurate potentials with mobility edges. With amplitude increments of the incommensurate potentials, the system can undergo a transition from a topological phase to a topologically trivial localized phase. Interestingly, we find that there are four mobility edges in the spectrum when the strength of the incommensurate potential is below a critical threshold, and a novel topologically nontrivial localized phase emerges in a certain region. We reveal this energy-dependent metal-insulator transition by applying several numerical diagnostic techniques, including the inverse participation ratio, the density of states and the Lyapunov exponent. Nowadays, precise control of the background potential and the $p$-wave superfluid can be realized in the ultracold atomic systems, we believe that these novel
mobility edges can be observed experimentally.

\end{abstract}

\maketitle

\section{Introduction}
\label{n1}
In recent years, considerable attention has been paid to the topological matters, including topological insulators (TIs)~\cite{K1,Zh2} and topological superconductors (TSCs)~\cite{K3,I4}. Among  various models, the one dimensional (1D) TSC, i.e., the spinless $p$-wave superconductor model studied originally by Kitaev~\cite{K3}, is an important and well known example. A key feature of the 1D TSC is that it hosts the zero-energy Majorana fermion states~\cite{S5,K6,Lu7}, which promise a platform for the error-free quantum computation since the information can be stored in the topologically protected Majorana states and the qubits are immune to the weakly disordered perturbation~\cite{P8}. However, if the time reversal symmetry of the 1D TSC system is broken by the presence of impurities~\cite{A9} or the strength of the disorder is strong enough, the stability of the topological phase can be significantly affected and a transition driven to the topologically trivial localized phase can occur.

The disorder effects of 1D TSC systems have been studied intensively. So far, most of the theoretical work for the Anderson localization in 1D
TSCs focuses on the random disorder~\cite{B10,G11,L12,B13} and the quasiperiodic disorder/incommensurate potential~\cite{14PRL,15PRB,liu,zhou,he,Gramsch,16A}. Ref.~\cite{14PRL} studies the interplay between the quasiperiodic disorder and superconductivity, and it leads to the topological phase transition from a topological superconducting phase to a topologically trivial localized phase when the strength of the incommensurate potential increases above a critical value. The same model is studied in Ref.~\cite{15PRB}, and a wide critical region in the parameter space is discovered, which is quite different from the Aubry-Andr\'{e} (AA) model~\cite{16A} where the wave-functions are critical only at the phase transition point.

However, none of these disorder models, both the random and the quasiperiodic, can host the mobility edge. A study about the interplay between the disorder with mobility edges and the $p$-wave superconducting pairing is still absent to the best of our knowledge. Here we introduce a class of 1D potentials~\cite{thou17,sarma18} with analytical expressions for the mobility edges, which enables us to study the interplay between the mobility edges and the $p$-wave superconducting pairing in a more controlled fashion. These deterministic potentials are neither random nor simply incommensurate, but rather slowly varying in real space. So we consider the 1D $p$-wave superconductor in these lattices, which is described by the following Hamiltonian
\begin{equation}\label{tb1}
    \hat H=\sum_{i=1}^{L-1}(-t \hat{c}_{i}^{\dag } \hat{c}_{i+1}+ \Delta \hat{c}_{i} \hat{c}_{i+1}+ H.c.)+\sum_{i=1}^{L}V_{i} \hat{n}_{i},
\end{equation}
where $\hat{c}^\dagger_i$ ($\hat{c}_i$) is the fermion creation
(annihilation) operator, $\hat{n}_i=\hat{c}^\dagger_{i}\hat{c}_{i}$ is the particle number operator,
and $L$ is the total number of sites. Here the nearest-neighbor
hopping amplitude $t$ and the $p$-wave pairing amplitude $\Delta$
are real constants, and $V_{i}=V\cos(2\pi\beta{i^{v}}+\phi)$ is the slowly varying incommensurate potential with $0<v<1$ and $V>0$ being the strength of the incommensurate potentials. A typical choice for parameters is $\beta=(\sqrt{5}-1)/2$, $\phi = 0$ and $v=0.4$. For computational convenience, $t = 1$ is set as the energy unit.

When $\Delta = 0$ and $v = 1$, this model reduces to the AA model, and the system can undergo a metal-insulator transition at $V=2$. When $\Delta = 0$ and $0<v<1$, Eq.(\ref{tb1}) describes a model with slowly varying incommensurate potentials~\cite{sarma18}. It is well known that this model has two mobility edges when $V<2$, i.e., all wave-functions with eigenenergy in $[V-2,2-V]$ are extended and otherwise localized. When $V>2$, all wave-functions are localized as in the AA model. When $\beta=0$ such that $V_{i}$ becomes a constant $V$, Eq.(\ref{tb1}) describes Kitaev's $p$-wave superconductor model, and the system can undergo a topological phase transition at $V=2$. When $\Delta \neq 0$ and $v = 1$, Eq.(\ref{tb1}) describes the 1D $p$-wave superconductor in incommensurate potentials. By applying this model, Ref.~\cite{14PRL} determines the phase transition point $V'= 2 + 2 \Delta$ both numerically and analytically, and Ref.~\cite{15PRB} demonstrates that wave-functions in the parameter space between $V''= 2 - 2 \Delta$ and $V'= 2 + 2 \Delta$ are not extended but critical.

In this work, we study the situation for which $\Delta \neq 0$ and $0<v<1$, i.e., the interplay between the disorder with mobility edges and the $p$-wave superconducting pairing. The main questions that we are interested are: (1) how the slowly varying incommensurate potentials drive a 1D $p$-wave superconductor to undergo a transition from a topological phase to a trivial phase, (2) how localized properties (such as mobility edges) of this system change besides the topological transition.

The rest of the paper is organized as follows. In Sec.~\ref{n2}, we investigate the phase transition from a
topological phase to topologically trivial localized phase. In Sec.~\ref{n3}, we demonstrate the existence of the four mobility edges by numerically studying the inverse participation ratio of wave-functions, the density of states and the Lyapunov exponent. We conclude and discuss possible experimental observations  in Sec.~\ref{n4}.

\section{Phase Transition from Topological phase to topologically trivial localized phase}
\label{n2}

\begin{figure}
  \centering
  \includegraphics[width=0.5\textwidth]{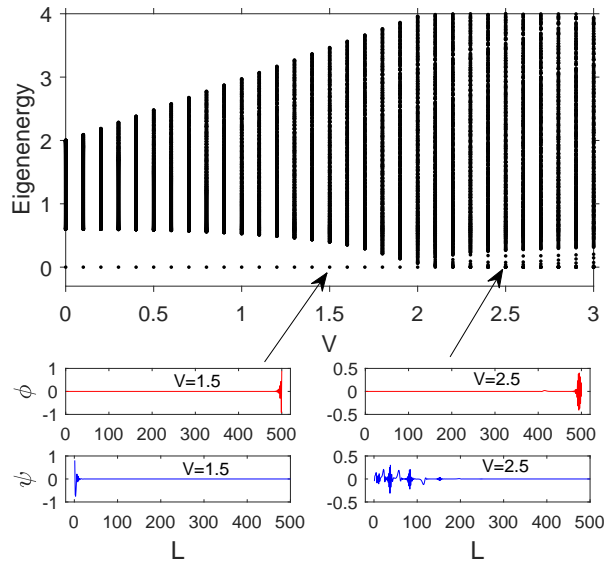}\\
  \caption{(Color online) The spectrum of the Hamiltonian in Eq.(\ref{tb1}) with $\Delta=0.3$ as a function of $V$ under the open boundary condition. Here the total number of sites is set as $L=500$. The spatial distributions of $\phi$ and $\psi$ for the lowest excitation with various $V$'s are shown in the lower figures. The lower left picture corresponds to the wave-functions of the zero energy mode, and the lower right picture corresponds to the wave-functions of the non-zero energy mode.}
  \label{001}
\end{figure}
\begin{figure}
  \centering
  \includegraphics[width=0.5\textwidth]{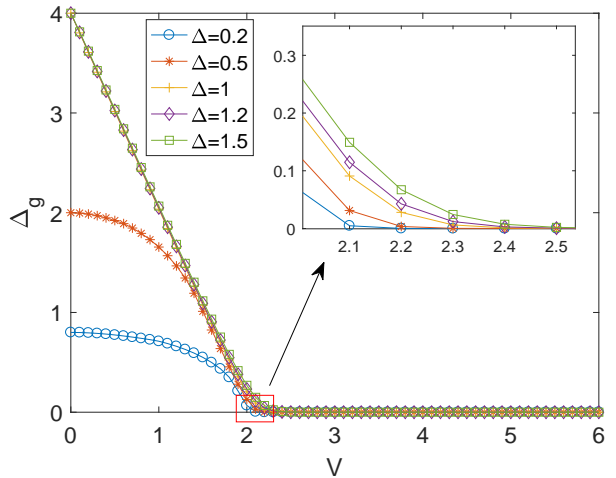}\\
  \caption{(Color online) $\Delta_g$ as a function of $V$ with various $\Delta$ under the periodic boundary condition. The total number of sites is set to $L=1000$. Here $\Delta_g$ is chosen to be twice of the lowest excitation energy. The inset is the blow up of of the changing trend of  $\Delta_g$ near $V_L=2$. We can clearly see that the energy gap $\Delta_g$ closes at various $V$'s and $\Delta$'s, which means the phase transition point is not a fixed value but in a range. }
  \label{002}
\end{figure}
The Hamiltonian~(\ref{tb1}) can be diagonalized by using the Bogoliubov-de
Gennes (BdG) transformation \cite{D19,L20}:
\begin{equation}
\hat{\chi} _{n}^{\dag } = \sum_{i=1}^{L}[u _{n,i} \hat{c}_{i}^{\dag } +
v _{n, i} \hat{c}_{i}], \label{quasi}
\end{equation}
where $L$ denotes the total number of sites, $n$ is the
energy level index, and
$u _{n, i}$, $v_{n, i}$ are the two-component wave-functions. Hence the
Hamiltonian is diagonalized as $H=\sum_{n=1}^{L}E _{n}(\hat{\chi}
_{n}^{\dag }\hat{\chi} _{n}-\frac{1}{2})$ where $E _{n}$ is the eigenenergy of the Hamiltonian.
The BdG equations can be expressed as
\begin{eqnarray}
 \left(
\begin{array}{cc}
\hat{m} & \hat{\Delta} \\
-\hat{\Delta} & -\hat{m}%
\end{array}
\right)
 \left(
\begin{array}{c}
u _{n} \\
v _{n}%
\end{array}%
\right) =
E _{n} \left(
\begin{array}{c}
u_n \\
v_{n}%
\end{array}
\right), \label{BDG}
\end{eqnarray}
where $ \hat{m}_{ij} = -t (\delta_{j,i+1} + \delta_{j,i-1}) +
V_{i}  \delta_{ji}$, $\hat{\Delta}_{ij} = - \Delta
(\delta_{j,i+1}-\delta_{j,i-1})$,
$u_n^T=(u_{n,1},\cdots,u_{n,L})$ and
$v_n^T=(v_{n,1},\cdots,v_{n,L})$.
It is widely known that the particle-hole symmetry $ \hat{\chi}_n (E _{n})=
\hat{\chi}_n^{\dagger} (-E _{n})$ is conserved in the BdG equtions.

By numerically solving Eq.(\ref{BDG}), we can get the spectrum of the system and the wave-functions $u_{n,j}$ and $v_{n,j}$.
In Fig.~\ref{001}, we show the spectrum when $\Delta=0.3$ under the open boundary conditions. It can be shown that there is a regime
with nonzero energy gaps and zero energy modes when $V\lesssim2$. Here the zero energy modes correspond to the Majorana edge states localized at the ends of 1D chain. When $V$ is above a certain value, there are neither obvious gaps separating the negative and positive parts of the spectrum nor zero energy modes. To show the Majorana edge states clearly,
we introduce $\gamma_i^A =
\hat{c}_{i}^{\dag }+ \hat{c}_{i}$ and $\gamma_i^B = (\hat{c}_{i}-
\hat{c}_{i}^{\dag })/i$,  where $\gamma^A$ and $\gamma^B$ are
two species of Majorana fermions, satisfying the relations
$(\gamma_i^\alpha)^{\dagger} = \gamma_i^\alpha$ and
$\{\gamma_i^\alpha, \gamma_i^\beta\} = 2
\delta_{ij} \delta_{\alpha \beta}$ with $\alpha$ and $\beta$
taking $A$ or $B$. Then the Bogoliubove quasi-particle operators can be rewritten as
\begin{eqnarray}
\hat{\chi} _{n}^{\dag } = \frac{1}{2}\sum_{i=1}^{L}[ \phi _{n, i}
\gamma_i^A  - i \psi _{n,i} \gamma_i^B ],
\end{eqnarray}
where $\phi _{n, i}=(u_{n, i}+ v_{n, i})$ and $ \psi _{n, i}=(u_{n, i} - v_{n, i})$.

To clearly show the difference between zero and non-zero energy modes, we plot the spatial distributions of $\phi$ and $\psi$ for the lowest excitation of the spectrum. When $V=1.5$, $\phi$ and $\psi$ of the zero energy modes are located at the right (left) end and decay very quickly away from the right (left) edge, as shown in Fig.~\ref{001}. Since there is no overlap between the amplitudes of $\phi$ and $\psi$, the zero energy modes split into two spatially separated Majorana edge states. However, when $V=2.5$ the amplitudes of $\phi$ and $\psi$ with the lowest excitation energy overlap together and are located within a finite range of the whole chain. This indicates the corresponding quasi-particle is a localized fermion which can not be split into two independent Majorana edge states. Therefore, these results demonstrate that the system can undergo a transition from a topological phase to a topologically trivial localized phase when the strength of the incommensurate potentials $V$ is increased to a certain level.

We now wonder if there exists a fixed value of $V$ to denote the phase transition point. In Fig.~\ref{002}, we plot the variation of energy gap $\Delta_g$ versus $V$ for different $\Delta$'s. The energy gap $\Delta_g$ vanishes near $V_L=2$, and the details can be found in the blow up of the $\Delta_g$ curve shown in the inset. Interestingly, the gap-closing points for different $\Delta$'s do not converge to a single point, hence the phase transition points spread over the region around $V_L=2$. An acceptable explanation for this phenomenon is that due to the slowly varying incommensurate potential $V_{i}=V\cos(2\pi\beta{i^{v}})$, its derivative is
\begin{equation}\label{tb2}
\frac{dV_{i}}{di}=-2V\pi\beta i^{v-1}\sin(2\pi\beta{i^{v}}).
\end{equation}
In the thermodynamic limit $ i \rightarrow \infty$, we have
\begin{equation}\label{tb3}
\lim_{i\rightarrow\infty} \mid\frac{dV_{i}}{di}\mid=-\lim_{i\rightarrow\infty}  2V\pi\beta \frac{\mid\sin(2\pi\beta{i^{v}})\mid}{i^{1-v}}=0,
\end{equation}
since $0<v<1$. Equivalently, $\lim_{i\rightarrow\infty}( V_{i+1} - V_{i})= 0$, which implies that the potential $V_{i}$ varies very slowly. This asymptotic property of ``being constant" of $V_{i}$ is similar to that of the chemical potential of Kitaev's $p$-wave model. It may explain why the phase transition points spread out near $V_L=2$ in our model.

\section{Mobility edges and topologically nontrivial localized phase}
\label{n3}
\begin{figure}
  \centering
  \includegraphics[width=0.5\textwidth]{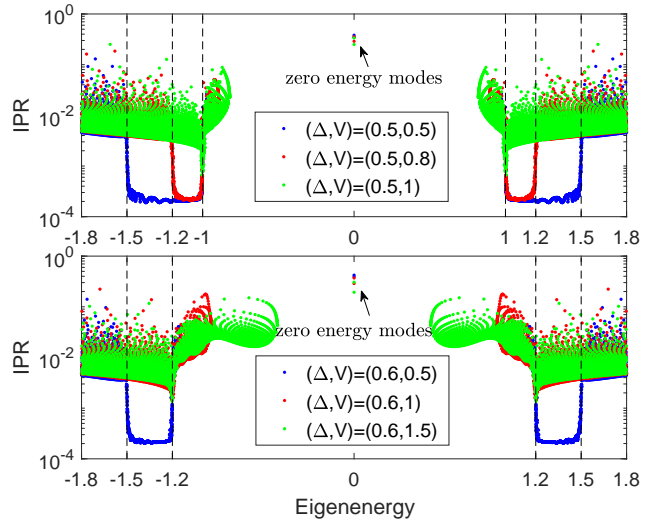}\\
  \caption{(Color online) The distribution of IPR as a function of eigenenergy for various
	$(\Delta,V)$. ``Black dotted lines" correspond to two turning points of IPR located at the mobility edges $E_{c1}=\pm (2- V)$ and $E_{c2}=\pm 2\Delta$ respectively. (a) When $(\Delta,V)= (0.5,0.5)$ and $(0.5,0.8)$, $E_{c1}=\pm 1.2, \pm 1.5 $ and $E_{c2}=\pm 2\Delta=\pm 1$ are located at the spectrum due to $V<2-2\Delta=1$, while when $(\Delta,V)= (0.5,1)$, the mobility edges disappear at the spectrum due to $V=2-2\Delta=1$.
	(b) When $(\Delta,V)= (0.6,0.5)$, $E_{c1}=\pm 1.5$ and $E_{c2}=\pm 1.2$ are located at the spectrum due to $V<2-2\Delta=0.8$, while when $(\Delta,V)= (0.6,1)$ and $(0.6,1.5)$, there are no mobility edges and all wave-functions are localized due to $V>2-2\Delta=0.8$, however, the zero energy modes still exist. Therefore, when the strength of the incommensurate potentials is less than the threshold $V_c = 2 - 2 \Delta$, there exist four mobility edges located at $E_{c1}=\pm (2- V)$ and $E_{c2}=\pm 2\Delta$ in the spectrum. The number of sites is set as $L=5000$ hereinafter in this paper.}
  \label{003}
\end{figure}
\begin{figure}
  \centering
  \includegraphics[width=0.5\textwidth]{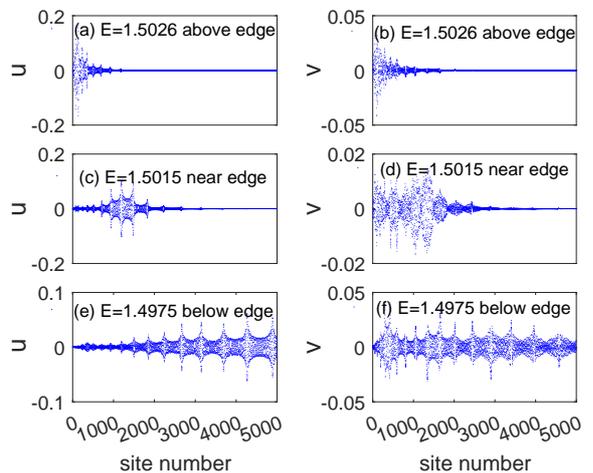}\\
  \caption{(Color online) Eigenstates $u$ and $v$ near the mobility edge $E_{c1}=1.5$, when $\Delta=0.5$ and $V=0.5$. Here we choose three typical eigenenergies (with four significant digits): high energy localized state above $E_{c1}$ ((a), (b)), critical state near $E_{c1}$ ((c), (d)), and low energy extended state below $E_{c1}$ ((e), (f)).}
  \label{004}
\end{figure}
\begin{figure}
  \centering
  \includegraphics[width=0.5\textwidth]{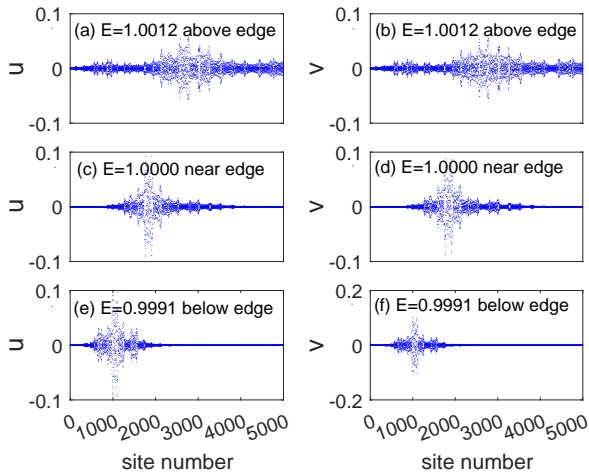}\\
  \caption{(Color online) Eigenstates $u$ and $v$ near the mobility edge $E_{c2}=1.0$, when $\Delta=0.5$ and $V=0.5$. Here we choose three typical eigenenergies (with four significant digits): high energy extended state above $E_{c2}$ ((a), (b)), critical state near $E_{c2}$ ((c), (d)), and low energy localized state below $E_{c2}$ ((e), (f)).}
  \label{005}
\end{figure}

Furthermore, to clarify the localized properties of this model we calculate the inverse participation ratio (IPR)~\cite{IPR21,IPR22,IPR23}, which is defined as
\begin{equation}
\text{IPR}_n =\sum_{j=1}^{L} (u_{n,j}^4 + v_{n,j}^4),
\end{equation}
for a normalized wave-function.
Here $n$ is the energy level index, and $u_{n,j}$, $
v_{n,j}$ are the solutions to BdG equations subject to the normalization condition $ \sum_i (u_{n,i}^2 + v_{n,i}^2)=1$.  The above definition can be thought of as an extension of IPR with $\Delta=0$.
It is well known that the IPR scales as $L^{-1}$ for an extended state. Hence it approaches $0$ in the thermodynamic limit, but is finite for a localized state.

Fig.~\ref{003} plots the IPR of the corresponding wave-functions as a function of eigenenergy for various $(\Delta,V)$. We find that as the eigenenergy varies,
the IPR suddenly jumps from the order of magnitude $10^{-2}$ (a typical value for the localized states) to $10^{-4}$ (a typical value for the extended states) or inversely at specific energies.
This jumping phenomenon suggests that there exist mobility edges in the energy spectrum. We did calculations for various $(\Delta,V)$ and found that these mobility edges are exactly located at $E_{c1}=\pm (2- V)$ and $E_{c2}=\pm 2\Delta$ respectively. For the mobility edges to exist there is an implicit condition that $2-V>2\Delta$. 
In Fig.~\ref{003}, it is clearly shown that when the strength of the slowly varying incommensurate potentials is larger than the threshold $V_c = 2 - 2 \Delta$, there are no mobility edges in the spectrum.

Remarkably, when $V>V_c$, the IPR of all wave-functions are about $10^{-2}$, and none of them appears around $10^{-4}$, as shown in Fig.~\ref{003}. Hence all wave-functions are localized in this situation. However, if $V_c<V<V_L$, there exists a region $[V_c,V_L]$ in which the energy gap does not close and the zero energy modes still exist as demonstrated in Fig.~\ref{003} and Fig.~\ref{001}. For the case $(\Delta,V)=(0.3,1.5)$ shown in Fig.~\ref{001}, although all wave-functions are localized due to $1.5>2-2\Delta=1.4$, $\phi$ and $\psi$ with the lowest excitation still split into two spatially separated Majorana edge states, therefore a novel topologically nontrivial localized phase emerges here. We choose different sets of parameters to ensure that this novel phase indeed exists by numerics.

Fig.~\ref{004} and Fig.~\ref{005} present the eigenstates corresponding to three different eigenenergies with $(\Delta,V)= (0.5,0.5)$. In Fig.~\ref{004}, the wave-function is localized (Fig.~\ref{004}(a) and (b)), critical (Fig.~\ref{004}(c) and (d)) and extended (Fig.~\ref{004}(e) and (f)), when the corresponding eigenenergy is above, near, and below the mobility edge $E_{c1}=2- V=1.5$ respectively. In Fig.~\ref{005}, in contrast, the wave-function is extended (Fig.~\ref{005}(a) and (b)), critical (Fig.~\ref{005}(c) and (d)) and localized (Fig.~\ref{005}(e) and (f)), when the corresponding eigenenergy is above, near, and below the mobility edge $E_{c2}=2\Delta=1$ respectively.


To strengthen our findings, we also calculate the density of states (DOS) $D(E)$ and the Lyapunov exponent $\gamma(E)$ of this system, which are defined as~\cite{sarma18}
\begin{align}
D(E) &=\sum_{n=1}^{L} \delta(E-E_n),\notag\\
\gamma(E_n) &=\frac{1}{L-1}\sum_{n\neq m}^{L}\ln| E_n-E_m|.
\end{align}
Here $E_n$ is the $n$-th eigenenergy. Since the Lyapunov exponent is the inverse of the localization length, then $\gamma=0$ for an extended state whereas $\gamma \neq0$ for a localized state. These two quantities are related to each other through the equation
\begin{equation}
\gamma(E) =\int dE'D(E') \ln|E- E'|.
\end{equation}

\begin{figure}
  \centering
  \includegraphics[width=0.5\textwidth]{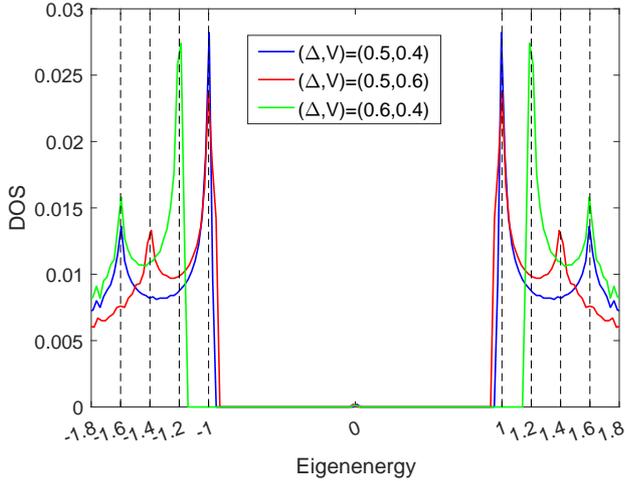}\\
  \caption{(Color online) DOS as a function of eigenenergy with three different sets of parameters $(\Delta,V)=(0.5,0.4)$, $(0.5,0.6)$ and $(0.6,0.4)$. Obviously a dramatic change occurs when the eigenenergy passes through the mobility edges $E_{c1}=\pm (2- V)$ and $E_{c2}=\pm 2\Delta$, which are in accordance with the IPR predictions.  }
  \label{006}
\end{figure}

\begin{figure}
  \centering
  \includegraphics[width=0.5\textwidth]{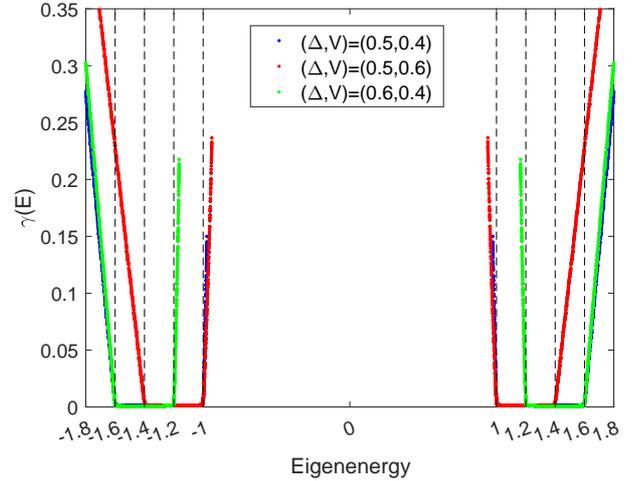}\\
 \caption{(Color online) The Lyapunov exponent $\gamma(E)$ vs eigenenergy with three different sets of parameters $(\Delta,V)=(0.5,0.4)$, $(0.5,0.6)$ and $(0.6,0.4)$. When the eigenenergy is located in the intervals $[V-2, -2\Delta]$ and $[2\Delta, 2 - V]$, $\gamma(E)\rightarrow 0$, indicating that the corresponding state is extended. Otherwise $\gamma(E)$ is finite, indicating that the corresponding state is localized. }
  \label{007}
\end{figure}

In Fig.~\ref{006} we present the behavior of DOS as a function of eigenenergy. Three different sets of parameters $(\Delta,V)=(0.5,0.4)$, $(0.5,0.6)$ and $(0.6,0.4)$ are chosen for not losing generality. The energy band consists of two subbands which are symmetric around $E=0$ due to the particle-hole symmetry. Obviously the DOS in our model is singular while crossing the mobility edge, and the change of the nature of the eigenstates can be reflected by the singularity of the DOS~\cite{thou17,sarma18}. Therefore two sharp peaks in both subbands shown in Fig.~\ref{006} indicate the extended state-localized state transition corresponding to two mobility edges located at $E_{c1}=\pm (2- V)$ and $E_{c2}=\pm 2\Delta$. In Fig.~\ref{007} we plot the Lyapunov exponent by plugging in the same sets of parameters as in Fig.~\ref{006}. It also exhibits a singular behavior at the mobility edge. The implications from the numerical results are in excellent agreement with those from the IPR and DOS. We also try other sets of parameters and obtain the same results as expected.

Another interesting subject is the specific form of the critical behavior of the Lyapunov exponent at the mobility edge. In the localized regions of energy spectrum, we have
\begin{equation}
\gamma(E) \sim | E- E'|^\theta.
\end{equation}
 Similarly, the density of states at the mobility edge behaves like
\begin{equation}
D(E) \sim | E- E'|^{-\delta}.
\end{equation}
The critical exponents $\theta$ and $\delta$ are related by the equation
\begin{equation}
\theta + \delta =1.
\end{equation}
In Fig.~\ref{007}, the singular behaviors of $\gamma(E)$ are identified to be linear with $E$ in the localized region, indicating that $\theta=1$, and $\delta=0$ accordingly. These results are the same as those of the single-particle model~\cite{sarma18}, and we find that the parameters $V$, $\Delta$, $\beta$ and $v$ are all irrelevant with regard to the critical exponents $\theta$ and $\delta$. In addition, by varying the parameters, we also find that the four mobility edges depends on $V$ and $\Delta$ but are irrelevant to $\beta$ and $v$.


\section{Conclusions}
\label{n4}
In summary, we study the interplay between the disorder with mobility edges and the $p$-wave superconducting pairing. With regard to the questions raised in the introduction, we find following interesting features of this model.

(1)  Increasing the strength $V$ of slowly varying incommensurate potentials can destroy the topological SC phase and drive
the system into a topologically trivial localized phase. The phase transition point occurs not at a fixed value of $V$ but in a region around $V_L=2$.

(2) There exist four mobility edges located at $E_{c1}=\pm (2- V)$ and $E_{c2}=\pm 2\Delta$ in the spectrum when the strength of the incommensurate potentials is less than a threshold $V_c = 2 - 2 \Delta$, otherwise all wave-functions are localized. Hence there is a region marking the topologically nontrivial localized phase between $V_c$ and $V_L$. To the best of our knowledge it has never been proposed in the 1D TSC system yet. We verified our predictions by utilizing several typical numerical techniques, and all results are consist with one another. We believe that the interesting features of this model will shed light on a wide range of topological and disordered systems.

Finally, we would like to point out that Anderson localization in disordered systems has been studied extensively in ultracold atomic experiments, both for the speckle disorder case~\cite{B24} and the quasiperiodic disorder case~\cite{R25} in controlled artificial method. Experimentally determining the mobility edge trajectory have been realized in a speckle disorder system with sufficiently high energy resolution~\cite{D26,J27,S28}. It is also possible to induce directly superfluid $p$-wave pairing by using a Raman laser in proximity to a molecular BEC~\cite{Jiang,Nascimbene}. These significant advances in ultracold atomic systems provide a potential way to experimentally study the interplay between mobility edges and the $p$-wave superconductor(superfluid). Thus we expect that these novel features including mobility edges and the topologically nontrivial localized phase discovered in this model can be
realized experimentally in the ultracold atomic system.
\begin{acknowledgments}
G. H. thanks the support from the NSF of China (Grant No.
11674051).
\end{acknowledgments}



\begin{thebibliography}{10}
\bibitem{K1} M. Z Hassan and C. L. Kane, Rev. Mod. Phys. {\bf 82}, 3045 (2010).
\bibitem{Zh2} X. L. Qi and S.-C. Zhang, Rev. Mod. Phys. {\bf 83}, 1057 (2011).
\bibitem{K3} A. Y. Kitaev, Phys. Usp. {\bf 44}, 131 (2001).
\bibitem{I4} D. A. Ivanov, Phys. Rev. Lett. {\bf 86}, 268 (2001).

\bibitem{S5} M. Stone and S.-B. Chung, Phys. Rev. B {\bf 73}, 014505 (2006).

\bibitem{K6} L. Fu and C. L. Kane, Phys. Rev. Lett. {\bf 100}, 096407 (2008).

\bibitem{Lu7} R. M. Lutchyn, J. D. Sau, and S. Das Sarma, Phys. Rev. Lett.
{\bf 105}, 077001 (2010).
\bibitem{P8} A. C. Potter and P. A. Lee, Phys. Rev. Lett. {\bf 105}, 227003
(2010).
\bibitem{A9} A. Altland and M. R. Zirnbauer,
Phys. Rev. B {\bf 55}, 1142 (1997).
\bibitem{B10} P. W. Brouwer, A. Furusaki, I. A. Gruzberg, and C. Mudry,
Phys. Rev. Lett. {\bf 85}, 1064 (2000).
\bibitem{G11} I. A. Gruzberg, N. Read, and S. Vishveshwara, Phys. Rev. B
{\bf 71}, 245124 (2005).
\bibitem{L12} A. Lobos, R. Lutchyn, and S. Das Sarma, Phys. Rev. Lett. {\bf 109}, 146403 (2012).
\bibitem{B13} P. W. Brouwer, M. Duckheim, A. Romita, and F. von Oppen,
hys. Rev. Lett. {\bf 107}, 196804 (2011).
\bibitem{14PRL} X. Cai, L.-J. Lang, S. Chen, and Y. Wang,
Phys. Rev. Lett. {\bf 110}, 176403 (2013).
\bibitem{15PRB} J. Wang, X.-J. Liu, G. Xianlong, and H. Hu,
Phys. Rev. B {\bf 93}, 104504 (2016).
\bibitem{liu} T. Liu, P. Wang, and G. Xianlong, arxiv:1609.06939 (2016).
\bibitem{zhou} L. Zhou, H. Pu, and W. Zhang, Phys. Rev. A {\bf 87}, 023625 (2013).
\bibitem{he} K. He, I. I. Satija, C. W. Clark, A. M. Rey, and M. Rigol, Phys. Rev. A {\bf 85}, 013617 (2012).
\bibitem{Gramsch} C. Gramsch and M. Rigol, Phys. Rev. A {\bf 86}, 053615 (2012).
\bibitem{16A}S. Aubry and G. Andr\'{e}, Ann. Israel Phys. Soc. {\bf 3}, 133 (1980).
\bibitem{thou17}D. J. Thouless,
Phys. Rev. Lett. {\bf 61}, 2141 (1988).
\bibitem{sarma18}S. Das Sarma, S. He, and X. C. Xia,
Phys. Rev. Lett. {\bf 61}, 2144 (1988); Phys. Rev. B {\bf 41}, 5544 (1990).
\bibitem{D19} P. G. de Gennes, {\it Superconductivity of Metals and
Alloys} (Benjamin, New York, 1966).
\bibitem{L20}  E. Lieb, T. Schultz, and D. Mattis, Ann.Phys.
(N.Y.) \textbf{16}, 407 (1961).
\bibitem{IPR21} D. J. Thouless, Phys. Rep. {\bf 13}, 93 (1974).
\bibitem{IPR22} M. Kohmoto,
Phys. Rev. Lett {\bf 51}, 1198 (1983).
\bibitem{IPR23}  M. Schreiber, J. Phys. C {\bf 18}, 2493 (1985); Y. Hashimoto, K. Niizeki, and
Y. Okabe, J. Phys. A {\bf 25}, 5211 (1992).
\bibitem {B24}J. Billy, V. Josse, Z. Zuo, A. Bernard, B. Hambrecht,
P. Lugan, D. Cl\'{e}ment, L. Sanchez-Palencia, P. Bouyer, and A. Aspect, Nature {\bf 453}, 891 (2008).
\bibitem {R25}G. Roati, C. D'Errico, L. Fallani, M. Fattori,
C. Fort, M. Zaccanti, G. Modugno, M. Modugno, and M. Inguscio,
Nature {\bf 453}, 895 (2008).
\bibitem{D26} S. S. Kondov, W. R. McGehee, J. J. Zirbel, and B. DeMarco, Science \textbf{334}, 66 (2011).
\bibitem{J27} F. Jendrzejewski, A. Bernard, K. Muller, P. Cheinet, V. Josse, M. Piraud, L. Pezz\'e, L. Sanchez-Palencia, A. Aspect, and P. Bouyer, Nat.~Phys. \textbf{8}, 398 (2012).
\bibitem{S28} G. Semeghini, M. Landini, P. Castilho, S. Roy, G. Spagnolli,
A. Trenkwalder, M. Fattori, M. Inguscio, and G. Modugno, Nat.~Phys. \textbf{11}, 554 (2015).
\bibitem{Jiang} L. Jiang, T. Kitagawa, J. Alicea, A. R. Akhmerov, D. Pekker, G. Refael, J. I. Cirac,
E. Demler, M. D. Lukin, and P. Zoller, Phys. Rev. Lett. \textbf{106}, 220402 (2011).
\bibitem{Nascimbene} S. Nascimbene, J. Phys. B \textbf{46}, 134005 (2013).
\end{thebibliography}
\end{document}